\documentclass[twocolumn,secnumarabic,amssymb, nobibnotes, aps, prl]{revtex4-1}

\usepackage{amsmath}
\usepackage{graphicx}
\usepackage{epstopdf}
\usepackage{upgreek}

\begin{document}

\title{Demonstration of single-shot picosecond time-resolved MeV electron  imaging using a compact permanent magnet quadrupole based lens}

\author{D. Cesar}
\affiliation{Department of Physics and Astronomy, UCLA, Los Angeles, California 90095, USA}

\author{J. Maxson}
\affiliation{Department of Physics and Astronomy, UCLA, Los Angeles, California 90095, USA}

\author{P. Musumeci}
\affiliation{Department of Physics and Astronomy, UCLA, Los Angeles, California 90095, USA}

\author{Y. Sun}
\affiliation{Department of Physics and Astronomy, UCLA, Los Angeles, California 90095, USA}

\author{J. Harrison}
\affiliation{Department of Electrical Engineering, UCLA, Los Angeles, California 90095, USA}

\author{P. Frigola}
\affiliation{RadiaBeam Technologies, Santa Monica, California, USA}

\author{F. H. O'Shea}
\affiliation{RadiaBeam Technologies, Santa Monica, California, USA}

\author{H. To}
\affiliation{RadiaBeam Technologies, Santa Monica, California, USA}

\author{D. Alesini}
\affiliation{INFN-LNF, Via E. Fermi, 40-00044 Frascati, Rome, Italy}

\author{R. K. Li}
\affiliation{SLAC National Accelerator Laboratory, Menlo Park, CA, 94025, USA}

\date{today}

\begin{abstract}
We present the results of an experiment where a short focal length ($\sim$ 1.3~cm) permanent magnet electron lens is used to image  micron-size features of a metal sample in a single shot, using an ultrahigh brightness ps-long 4 MeV electron beam from a radiofrequency photoinjector. Magnification ratios in excess of 30x were obtained using a triplet of compact, small gap (3.5 mm), Halbach-style permanent magnet quadrupoles with nearly 600 T/m field gradients. These results pave the way towards single shot time-resolved electron microscopy and open new opportunities in the applications of high brightness electron beams.
\end{abstract}

\pacs{06.60.Jn}

\maketitle


Transmission electron microscopy (TEM) is one of the primary tools for materials characterization, with many scientific and industrial applications. One of the recent trends in TEM development is the quest for in-situ dynamic imaging in which a sequence of micrographs are captured in a time-resolved mode while the sample under study is undergoing some sort of microscopic rearrangement \cite{King:JAP,Zewail:uem,DOE:workshop}.

Improving the temporal resolution of TEMs to ultrafast time scales presents significant challenges. In order to substantially decrease the image acquisition time it is necessary to increase the peak current by many orders of magnitude. But at large currents the temporal resolution and transverse coherence rapidly degrade due to Coulomb interactions between the beam electrons \cite{Reed:spacecharge}. Thus, state-of-the-art single shot TEM systems have been limited to 10 nm 10 ns spatio-temporal resolution\cite{dtem,Browning:DTEM}. The only known remedy has been to reduce the number of charged particles per pulse and integrate over many millions of shots in order to collect a single picture \cite{caltechuem}. This technique has produced a variety of scientific results \cite{PINEM}, but it is restricted to fully reversible processes.

Single-shot picosecond transmission electron microscopy (SPTEM) would fill an unmet need in the TEM community to image irreversible dynamical motion at nm-ps spatio-temporal scales enabling real time study of the dynamics of many technologically and scientifically relevant microscopic processes, such as phase transitions and dislocation motion \cite{Lagrange:defectdynamics}. One path to SPTEM requires replacing the 100 keV typical of conventional TEM's with MeV electrons in order to take advantage of the relativistic suppression of the space-charge effects. This solution, discussed in detail in \cite{Li:PRApplied} involves a ground-up redesign in the microscope architecture, starting with the highest peak brightness source of relativistic electrons available to date, the radiofrequency (RF) photoinjector.

RF photoguns have played a central role in the development of the high brightness beams used in XFELs \cite{Emma:LCLS}. By combining the high current densities available in photoemission with the extremely high fields of a standing wave RF cavity, the RF photoinjector has already demonstrated the capability of generating MeV electron beams bright enough to capture single-shot diffraction patterns with a shutter speed of less than 100 fs \cite{Musumeci:APL,SLAC:MeVUED,Murooka:MeVUED,REGAE}.

One of the main challenges for SPTEM comes from the fact that the high electron energy, which conveniently limits the influence of Coloumb self-fields, comes at the cost of increased magnetic rigidity. High voltage (1-3 MeV) electron microscopes were, until the advent of aberration correction, one of the main candidates for improving the spatial resolution in TEM to atomic level \cite{Spence:book}. These machines were overburdened by large and expensive round solenoid lenses weighing up to several tons. The unfavorable scaling of solenoid focusing power as the inverse square of the electron energy poses a practical limit to the development of time-resolved electron microscopy \cite{Dao:MeVUEM,Yang:MeVUEM} and calls for the introduction of very strong magnetic lenses and/or of novel focusing elements.

Our approach borrows from experience in the field of advanced accelerators and involves the use of permanent magnet quadrupole (PMQ) lenses for imaging with relativistic electrons. PMQ triplets provide a compact short-focal-length lens for use by Inverse Compton Scattering sources \cite{Lim:PMQ} and advanced accelerator applications \cite{Sears:PMQ}.

In this paper we report on using a ps-long 4 MeV electron beam from an RF photoinjector and a strong compact PMQ-based lens with a focal length of $\sim$ 1.3~cm to obtain single-shot micrographs with $\mu$m-scale spatial resolution. The quadrupoles used in our experiment were measured to have field gradients of nearly 600 T/m, which to our knowledge set a new record for the strongest quadrupoles ever built. Magnification factors larger than 30x have been achieved. These results represent the first example of single shot ps-time resolved transmission electron microscopy.

The experiment was performed at the UCLA Pegasus Laboratory \cite{Musumeci:blowout} where a 1.6 cell S-band RF gun, fabricated using a brazeless clamped design \cite{Alesini:PegasusRFgun} is used to generate a high brightness electron beam. In order to maximize image sharpness, the photoinjector is operated in an ultra-low emittance configuration in which the laser spot on the cathode is minimized ($8 \times 12~\upmu$m). This is achieved by illuminating the photocathode from a $72^{\circ}$ port located in the first cell of the RF cavity, which allows the use of a high power final focus lens ($f$=17.5~cm). The small source size enables minimization of the initial phase space area, which is preserved during transport because the beam rapidly expands transversely into a uniformly filled ellipsoid \cite{Li:nmemittance}.

The beam is transported to the microscope sample plane located 3.7 m from the cathode using a two solenoid condenser which provides flexibility in choosing sample illumination. The tranverse beam parameters are characterized by inserting a thin (20~$\mu$m) YAG screen located shortly before the sample plane. The screen is imaged by an in-vacuum optical microscope objective with a 1~$\mu$m spatial resolution limited by a narrow depth of focus. On this screen the rms spot can be made as small as $3~\mu$m, with a normalized emittance (measured by scanning the solenoid current) of 5~nm, for a 20~fC beam. For larger beam charges (up to 100 fC), as employed in the experiments, the normalized emittance is measured below 20 nm in agreement with simulations performed using the General Particle Tracer (GPT) code \cite{gpt}. The electron beam duration was measured to be $0.9\pm0.15$ps (rms) using an x-band deflecting cavity operated as a streak camera located shortly after the microscope \cite{England:Deflector}.

\begin{figure}[!htb]
   \centering
   \includegraphics[width=.45\textwidth]{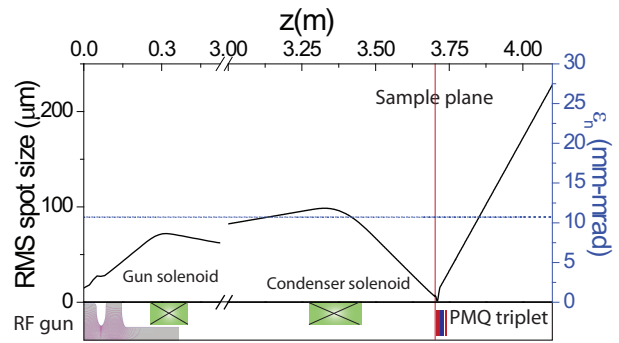}
   \caption{Schematic of the MeV TEM Pegasus beamline. The evolution of the RMS spot size and normalized emittance $\epsilon_n$ along the beamline from a GPT simulation for a 50 fC beam charge are also reported. Note the axis break.}
   \label{setup}
\end{figure}

A set of PMQs is placed just after the sample, designed to form an image of the sample 41 cm downstream. Obtaining imaging and equal magnification in both transverse dimensions requires a minimum of 3 quadrupoles. Solutions with 4 or 5 quadrupole magnets offer more flexibility at the expense of tighter alignment tolerances. To simplify machining and alignment, the three PMQs for this experiment were built with the same inner (3.5 mm) and outer (7 mm) diameter. The yoke lengths are 6 mm for the first two quadrupoles of the triplet and 3 mm for the last one respectively.

\begin{figure}[!htb]
   \centering
   \includegraphics[width=.45\textwidth]{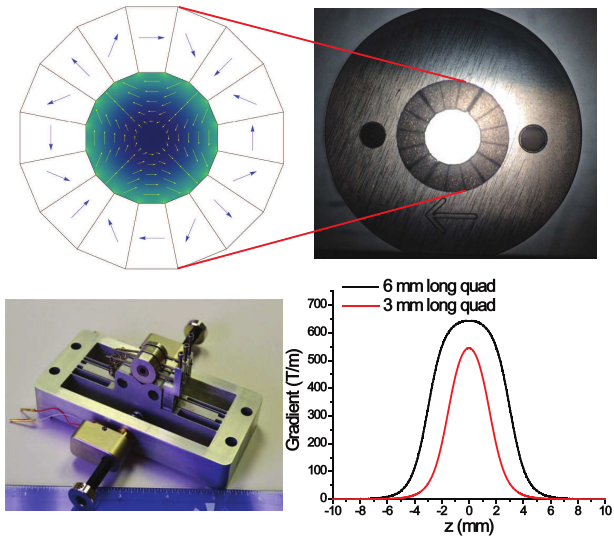}
   \caption{ a) 3d magnetic field map for an ideal Halbach PMQ, b) Photo of single PMQ, c) Picture of the PMQ triplet setup, c) Magnetic field profile of the PMQs from 3D Radia simulations.}
   \label{PMQ_figure}
\end{figure}

Each PMQ is made up of a 16 sector Halbach-style array of grade N35SH NbFeB wedges \cite{Halbach:PMQ} wire electrical discharge machined and assembled inside a precision machined aluminum “keeper” (see Fig. \ref{PMQ_figure}).
The 3D magnetostatic solver Radia \cite{RADIA} is used to compute the effective magnetic lengths (6.2 mm and 3.6 mm) and peak gradients (597 T/m and 495 T/m) for the long and short quadrupoles respectively. Due to the small aperture size it was challenging to obtain accurate Hall probe measurements of the field profile. A summary of the parameters of the PMQ triplet as well as the results of the magnetic measurements is shown in Table \ref{pmq_parameters}. A vibrating wire technique was used to measure the integrated field gradients \cite{Wolf:vibratingwire} which, within the relatively large error due to the calibration uncertainties, were found in good agreement with the expected values.

\begin{table*}[h!t]
    \caption{Parameters for the PMQ triplet. The reference position is measured from the sample plane.}
\begin{ruledtabular}
\begin{tabular}{lcccc}
      & \textbf{Design Gradient} 	& \textbf{Effective length}  & \textbf{Measured G $\times$ L} & \textbf{Reference design position} \\
\hline
\textbf{First quadrupole} & 597 T/m	& 6.16 mm & 3.3 $\pm$ 0.4 T & 5.25 mm \\
\textbf{Second quadrupole} & -597 T/m & 6.16 mm & 3.6 $\pm$ 0.5 T & 11.25 mm \\
\textbf{Third quadrupole} & 495 T/m	& 3.6 mm & 1.7 $\pm$ 0.2 T & 17.25 mm \\
\hline
\end{tabular}
\end{ruledtabular}
\label{pmq_parameters}
\end{table*}

A custom-design Al stage making use of flexures, displaced by linear piezo actuators, is used to adjust the longitudinal position of individual PMQs. The vibrating wire technique was used to determine the offsets and pre-align the magnets on the stage. The range of motion in the flexure-based stage (about 0.75 mm) allows achieving an imaging condition for a relatively wide range of input beam energies, i.e. between 3.5 and 4.75 MeV.

A picture of one of the assembled PMQs as well as the full triplet setup is shown in Fig. \ref{PMQ_figure}. The total weight of the PMQs and the flexure stage is less than 2 pounds. The wedge magnetization orientations as well as the magnetic field in the central plane of the quadrupole are also displayed.  Field maps for each of the fabricated magnets (including the individual wedge dimensions and the gaps originating from manufacturing errors) have been obtained using 3D magnetostatic simulations.

\begin{figure}[!htb]
   \centering
   \includegraphics[width=.45\textwidth]{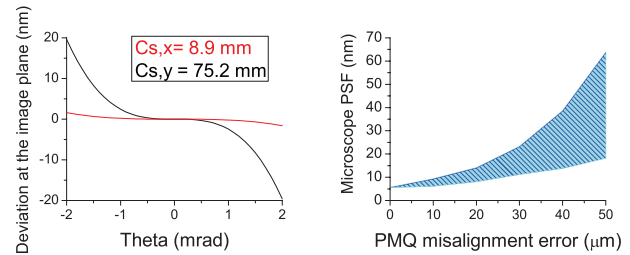}
   \caption{a) Aberrations for the PMQ triplet b) Tolerances to misalignment of the three-element lens. The shaded area is obtained calculating the rms size of the beam at the detector plane after tracking a very small source of electrons with 1 mrad divergence when each quadrupole is displaced in a random direction in the transverse plane by a fixed amount.}
   \label{aberrations}
\end{figure}

These field maps permit detailed simulations of the microscope column beam dynamics. We begin by solving a linear transport model of hard-edge quadrupoles to find out the beamline distances required to achieve an imaging condition with equal magnifications in $x$ and $y$ at the detector plane. We then refine the calculation by using the quadrupole gradient profile along the beamline axis, $z$. Finally, tracking the particle trajectories in the full PMQ triplet magnetic field maps was used to estimate the transverse tolerance to misalignment and the aberrations of the system. The results are shown in Fig. \ref{aberrations}. The calculated spherical aberrations for the manufactured PMQs are 8.9 mm and 75.2 mm in the horizontal and vertical plane respectively. The asymmetry could be reversed by using the vertically focusing quadrupole first instead of the horizontally focusing one. It was also found that for each quadrupole an angular misalignment of $\pm$ 10 $\mu$rad and a transverse displacement of 50 $\mu$m with respect to the central beam trajectory were required in order to avoid degradation of the image quality.



A 20$\pm 5~\mu$m thick Cu `UCLA' sample target was fabricated using lithographic techniques with varied feature sizes from $5~\mu$m to $100~\mu$m. The target was mounted on a 3 mm standard TEM holder and inserted in the beamline using a micrometer translation stage 500 $\mu$m from the front face of the first PMQ. A HeNe laser copropagating with the electron beam was used to align the sample to the axis of the PMQ triplet and the main beamline. The image was collected using a 100~$\mu$m thick YAG screen lens-coupled to a Princeton Instrument PIMAX III intensified camera. The point spread function (psf) of this phosphor screen-based imaging system (not to be confused with the psf of the microscope itself which depends on the magnification) is estimated to be ~50~$\mu$m rms, mostly attributable to the screen thickness.

\begin{figure}[!htb]
   \centering
   \includegraphics[width=.45\textwidth]{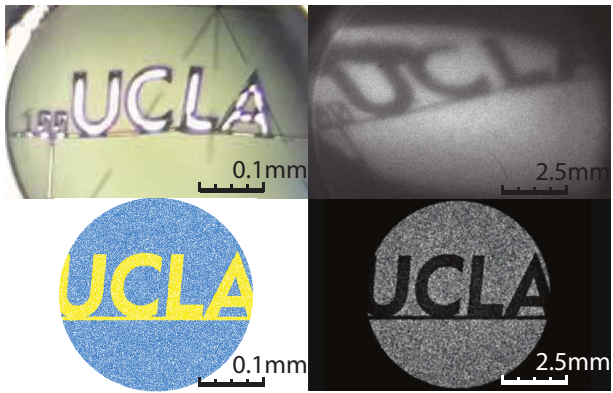}
   \caption{a)Optical and b)electron image of the nanofabricated `UCLA' target. c) Simulated distribution at the target. The color-coding indicates division between scattered and unscattered particles d) Simulated distribution at the image plane.}
   \label{UCLA_TEM}
\end{figure}

An optical image of the sample is shown next to a representative single-shot electron image of the sample in Fig. \ref{UCLA_TEM} (a,b). All of the sample features are clearly visible in the electron image, as is a contaminant which was introduced above the `U' during sample preparation. The skewness of the electron image is accentuated by alignment error so that the sample does not sit precisely perpendicular to or centered on the PMQ axis. The dimensions of the letters in the electron image can be used to compute a magnification of 32x and 25x in the horizontal and vertical plane respectively in fair agreement with the design magnification of 25x Fig. \ref{UCLA_TEM} (c,d). The astigmatism in the system is caused by the quadrupole placement and could be removed by fine-tuning the quadrupole positions.


A quantitative comparison of the simulated and recorded electron images requires a complete understanding of the electron imaging apparatus. Start to end simulations of the image formation process are performed taking into account multiple elastic and inelastic scattering of electrons inside the sample.  This is included in the particle tracking simulations by assigning an additional divergence and energy spread for particles that hit the metallic sample, accounting for the multiple elastic scattering and inelastic collisions, respectively \cite{Bethe:MultipleScattering,pdg}. The full simulation (Fig. \ref{UCLA_TEM}) shows that contrast is created when scattered electrons are clipped by the aperture of the magnets. Additional contrast is provided by the imperfect imaging of the lower energy electrons.

In both simulation and experiment, the highest resolution electron images are obtained at the maximum sample illumination flux, $n_e$=18 electrons/$\mu \text{m}^2$. Musumeci and Li \cite{Li:PRApplied} showed that as the charge density is increased beyond a certain optimum level, space charge effects and point-to-point scattering will cause image blurring. Given the relatively small magnification factor and large feature sizes, the impact of Coulomb scattering could not be measured in this experiment. Nevertheless, by varying the condenser lens strength, we were able to quantify the effect of changes in the illumination flux on the image sharpness.

\begin{figure}[!htb]
   \centering
   \includegraphics[width=.45\textwidth]{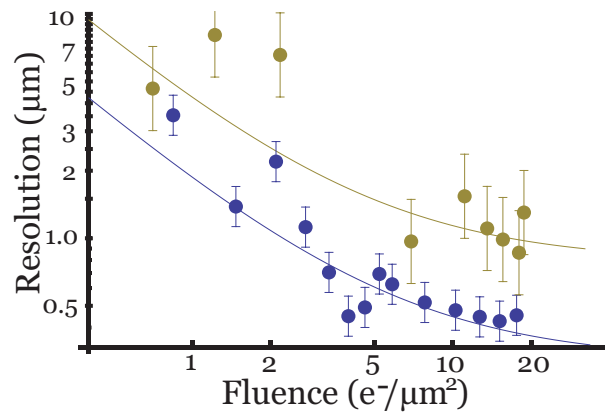}
   \caption{Microscope resolution as a function of charge for simulated (blue) and measured (gold) data. The error bars on the simulations are due to the random particle initialization. The solid lines show the $1/n_e$ scaling for the resolution.}
   \label{fig:rosecrit}
\end{figure}

In Fig. \ref{fig:rosecrit} we show the resolution in both experimental and simulated images quantified as the standard deviation of the centroid positions of the error-function fits to the line-outs taken along the edge of the `L' in the `UCLA' sample. The data points are obtained from a series of images captured with different condenser solenoid settings and beam charges (to vary $n_e$). GPT simulations are then performed using the measured illumination fluences. Both data and simulation show that the error on these positions (and therefore the image sharpness) improves as the fluence is increased. Assuming Poisson statistics for the signal we expect the spatial resolution in the image to scale as 1/$n_e$ according to the Rose criterion\cite{rose}. The inherent psf of the detector system further limits the spatial resolution. In order to quantify this, a gaussian blur of 20$~\mu$m (at the detector plane, and therefore 0.7 $\mu$m at the sample plane considering the 30x magnification) was taken into account when computing the simulated images. The main difference between the experimental and simulated curves is their asymptotic high-fluence-limit, which can be traced back to the differences between the simulated and real point spread functions discussed above. Fig. \ref{fig:rosecrit} serves to show that the resolution of the current microscope setup could be further enhanced by improving the detection system \cite{Minor:detectors}.

\begin{figure}[!htb]
   \centering
   \includegraphics[width=.45\textwidth]{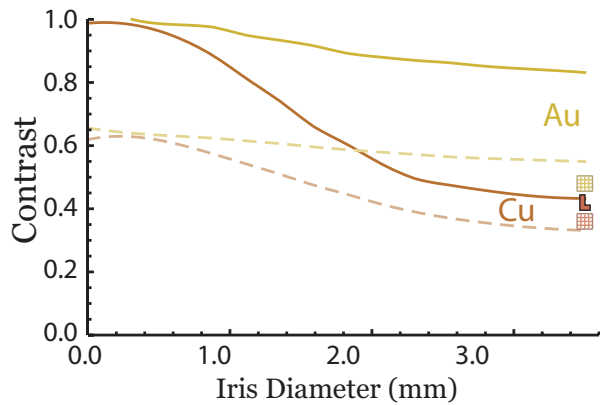}
   \caption{Simulated contrast as a function of the objective iris aperture for copper and gold targets having feature sizes similar (dashed) or well under (solid) the spatial resolution of the microscope. Also shown are the contrast of the 3 samples (UCLA target and Cu and Au TEM grids) observed experimentally using the PMQ gap as iris aperture.}
   \label{fig:contrast}
\end{figure}

For small image features, the resolution becomes intertwined with contrast such that understanding and improving contrast is a necessary component of a high magnification system. Contrast is defined from the image intensity as $(I_{max}-I_{min})/(I_{max}+I_{min})$. In Fig. \ref{fig:contrast} we show four simulated curves demonstrating the effect that adding an objective aperture would have on the image contrast. The two solid lines show the simulated contrast for copper and gold versions of the `UCLA' target. The effect of the iris size is more dramatic for the Cu since copper scatters the electrons less than gold. The rms angular and energy spread size of the gaussian distributions of the particles hitting the samples are  $\theta_{\text{Cu}}=0.1$, $\Delta E_{\text{Cu}}=29$~keV and $\theta_{\text{Au}}=0.2$, $\Delta E_{\text{Au}}=68$~keV for copper and gold respectively. The contrast of a copper target for an aperture equal to the gap between the PMQ magnets (3.5~mm) is 0.43, in close agreement with the 0.42 contrast obtained from analysis of the line profiles of the `L' in the electron images. Also shown are two dashed lines showing the results of simulations of image formation for objects having sizes similar to the psf of the detection system. In such cases the differences between gold and copper samples are significantly smaller as the contrast is dominated by the resolution, not by the sample scattering properties. These simulations can be compared to the measured contrast from $5~\mu$m bars on gold and copper TEM 2000 grids, shown in Fig \ref{fig:contrast} above and below the `L', respectively. Future single-shot time-resolved TEMs will require using an iris to increase the percentage of scattered electrons which are clipped. Diffraction contrast could also be obtained by positioning slits at the back focal plane(s) of the lens.

In conclusion these experiments demonstrate the first single shot, ps time-resolved electron images using high brightness relativistic beams from an RF photoinjector and the design and construction of a record-high gradient PMQ-based objective lens. This compact lens design can be used in subsequent magnification stages to approach the spatial resolution limits of the instrument. While the aberration coefficients of the quadrupole lens might seem high to conventional TEM microscopists, single shot picosecond TEM simulations indicate that the final spatial resolution after the addition of multiple magnification stages will be limited by space charge blurring \cite{Li:PRApplied}. Besides the reduction in cost, size and higher focusing power, quadrupole-based lenses might also offer an advantage over round lenses due to the smaller charge density that is obtained in elliptical cross-overs. The results reported in this paper validate the simulation models of the beam dynamics in the relativistic electron column and image formation process, paving the way towards the use of bright relativistic electron sources to achieve the long-range goal for single-shot time-resolved TEM of being able to follow defect dynamics in materials with 10~nm spatial resolution and ps temporal resolution.

This work was partially supported by DOE STTR grant No. DE-SC0013115 and National Science Foundation grant PHY-1415583. The authors want to acknowledge A. Murokh and G. Andonian for useful discussions.

\bibliographystyle{unsrt}

\end{document}